\RequirePackage{fix-cm}
\documentclass[smallextended]{svjour3}       
\smartqed  
\usepackage{graphicx}
\usepackage{dcolumn} 
\usepackage{bm} 
\usepackage{amssymb,amsmath} 
\usepackage{textcomp} 
\usepackage[utf8x]{inputenc}
\usepackage{braket}
\usepackage{amsmath}
\usepackage{placeins}
\usepackage{xcolor}
\usepackage{pdfpages}
\usepackage{soul}
\usepackage{hyperref}       
\usepackage{url}            
\usepackage{booktabs}       
\usepackage{todonotes} 
\usepackage{changebar}
\usepackage{changes}

\usepackage{amsfonts}
\usepackage{hyperref}
\begin{document}

\title{Order book regulatory impact on stock market quality: a multi-agent reinforcement learning perspective
}


\author{Johann Lussange$^{1}$
\and
Boris Gutkin$^{1,2}$
}


\institute{
$^{1}$ Group for Neural Theory, Laboratoire des Neurosciences Cognitives et Computationnelles, INSERM U960, D\'epartement des \'etudes Cognitives, \'ecole Normale Sup\'erieure, 29 rue d'Ulm, 75005, Paris, France. \\
$^{2}$ Center for Cognition and Decision Making, Department of Psychology, NU University Higher School of Economics, 8 Myasnitskaya st., 101000, Moscow, Russia. 
}

\date{Received: date / Accepted: date}

\maketitle

\begin{abstract}
Recent technological developments have changed the fundamental ways stock markets function, bringing regulatory instances to assess the benefits of these developments. In parallel, the ongoing machine learning revolution and its multiple applications to trading can now be used to design a next generation of financial models, and thereby explore the systemic complexity of financial stock markets in new ways. We here follow on a previous groundwork, where we designed and calibrated a novel agent-based model stock market simulator, where each agent autonomously learns to trade by reinforcement learning. In this Paper, we now study the predictions of this model from a regulator's perspective. In particular, we focus on how the market quality is impacted by smaller order book tick sizes, increasingly larger metaorders, and higher trading frequencies, respectively. Under our model assumptions, we find that the market quality benefits from the latter, but not from the other two trends. 

\keywords{agent-based model \and reinforcement learning \and order book \and multi-agent system \and stock markets \and regulation \and high-frequency trading \and tick \and decimalization \and metaorders \and machine learning}
\end{abstract}


\section*{Acknowledgements}We graciously acknowledge this work was supported by the HSE Basic Research Program and the Russian Academic Excellence Project ``5-100" and CNRS PRC nr. 151199, and received support from FrontCog ANR-17-EURE-0017. We also thank Michael Benzaquen and Bence T\'{o}th, from \textit{Capital Fund Management}, for their helpful insights and fruitful discussions.

\section{Introduction}
\label{SectionI}

Financial market economics has had a rich background in a bottom-up approach to systemic complexity~\cite{Sornette2014}. In this approach, one can especially mention agent-based modelling, which relies on multi-agent systems (MAS) to study and describe financial markets. These agent-based models (ABM) have been used to study, in particular, market regulatory impact~\cite{Boero2015,Furtado2016} and exogenous effects~\cite{Gualdi2015}, the impact of high-frequency trading~\cite{Wah2013,Aloud2014} and quantitative easing~\cite{Westerhoff2008}. Together with their cousin order book models, ABM have been used to study game theoretic~\cite{ErevRoth2014} and flow aspects of the law of supply and demand~\cite{Benzaquen2018} in order books~\cite{Huang2015}. 

\vspace{1mm}

Beyond all these particular studies, a main task for ABM and MAS has been to re-enact so-called \textit{stylised facts}, which are certain recurrent (or rather, \textit{lack} of) statistics found in financial markets, both across asset classes and over different time scales, that have been deemed universal~\cite{Bouchaud2018}. These are: non-gaussian price returns~\cite{Cristelli2014}, clustered price volatilities~\cite{Lipski2013}, and decaying price auto-correlations with time~\cite{Cont2005}. ABM have two major advantages in the way they can re-enact these stylized facts: firstly, they require fewer model assumptions (no gaussian distributions, no efficient market hypothesis~\cite{Fama1970,Bera2015}). Secondly, they naturally display the specific emergent phenomena proper to complex systems~\cite{Bouchaud2019}. 

\vspace{1mm}

A recurrent and historical critic of ABM addresses the challenges of modelling the agents constituting the model itself. However, the ongoing role played by machine learning and artificial intelligence in finance~\cite{Ganesh2019,Hu2019,Neuneier1997,Deng2017} changes the epistemological weight of this consideration. Notably, one should especially highlight how reinforcement learning~\cite{Charpentier2020,Silver2018}, with its numerous links to decision theory and the neurosciences~\cite{Eickhoff2018,Frydman2016,Lefebvre2017,Palminteri2015,Dayan2008}, could impact the use of ABM as statistical inference tools~\cite{Lussange2018}, not unlike what has been done with the recent, AI-augmented, order book models~\cite{Spooner2018,Biondo2019,Sirignano2019}. 

\vspace{1mm}

In a previous groundwork~\cite{Lussange2019}, we have described in details the design and parameter selection of a next generation stock market ABM, where each agent learns to forecast and trade by reinforcement learning in an autonomous fashion. This ABM has been cautiously calibrated to real financial data, namely the end-of-day stock prices and volumes of $642$ stocks from the London Stock Exchange, between years $2007$ and $2018$. In another previous work~\cite{Lussange2019b}, we also used this multi-agent reinforcement learning model to study agent learning and its mesoscale market impact. In this paper, we do not review all of the details of this model, but will simply describe in Section \ref{SectionII} its general architecture, together with the machine learning features of its agents. We shall also display some of its key results as supplementary material, in Fig. \ref{S1}-\ref{S4}. We will then show, in Section \ref{SectionIII}, how such a model can be used to study the impact of smaller order book tick sizes on market quality. Then, in Section \ref{SectionIV}, we will display some of the model's predictions concerning the impact of larger metaorders on the market microstructure. Finally, we shall study how market quality can benefit from higher frequency trading in Section \ref{SectionV}, and present a brief conclusion of all these results in Section \ref{SectionVI}.

\section{Groundwork}
\label{SectionII}

\begin{figure}[!htbp]
\begin{centering}
\includegraphics[scale=0.1]{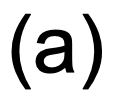} 
\includegraphics[scale=0.15]{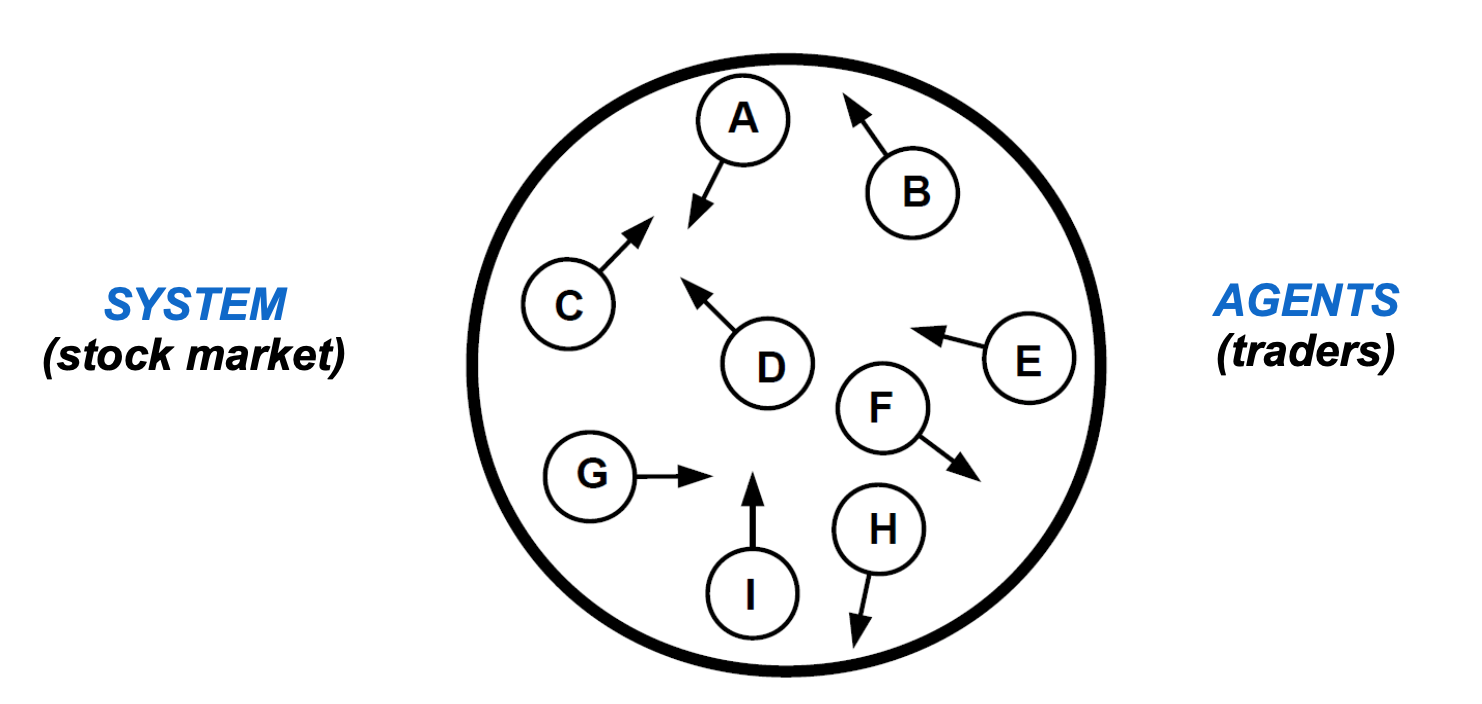}\\
\includegraphics[scale=0.1]{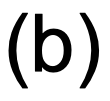}
\includegraphics[scale=0.15]{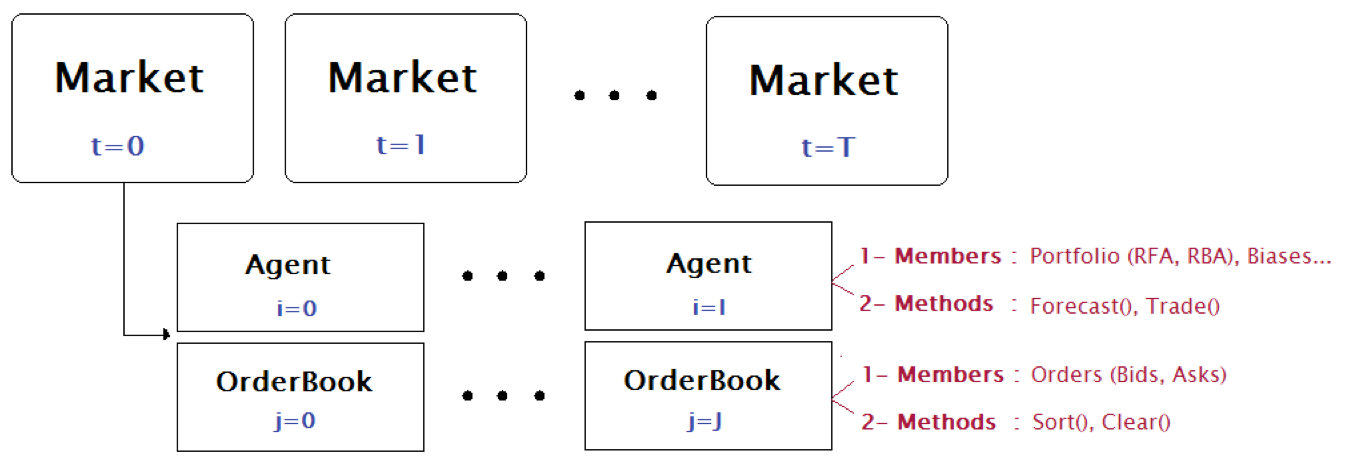}
\caption{\label{A1} (a) Diagram of the MAS stock market simulator for each stock $j$; (b) diagram of the $I$ agents and $J$ order books process over the simulations $T$ time steps.}
\end{centering}
\end{figure}

The ABM model of this groundwork~\cite{Lussange2019,Lussange2019b} relies on a collection of $I$ autonomous reinforcement learning agents, which trade a set of $J$ stocks over a simulation time of $T$ time steps, and over a number $S$ of simulations, as shown in the sum up diagram of Fig. \ref{A1}. At each time step $t$ of the simulation, the agents learn to better forecast future stock prices at their own investment horizon, and then may (or may not) send a trading order to a centralized, double-auction order book. At time $t=0$, each agent $i$ is first initialized with a given portfolio consisting of risk-free assets (e.g. a bank account) of value $A_{bonds}^{i}(t)$, and a number $Q^{i,j}(t)$ of stocks $j$, with a value $A_{equity}^{i}(t)=\sum_{j=0}^J Q^{i,j}(t)P^{j}(t)$. The stock price is initialized as $P^{j}(t=0)=\pounds 100$. A central challenge of these agents, just as for real traders and portfolio managers, is to derive an accurate pricing of the traded assets at their own investment horizon. Similarly to other models~\cite{Chiarella2007,Franke2011}, the model at time $t=0$ first generates a given time series $\mathcal{T}^{j}(t)$ corresponding to the fundamental values of each stock $j$. These fundamental values are not fully known by the agents, who compute an approximation of $\mathcal{T}^{j}(t)$ according to a proprietary rule $\kappa ^{i,j} [ \mathcal{T}^{j}(t) ]=\mathcal{B}^{i,j}(t)$ of cointegration~\cite{Murray1994}. The asset pricing process autonomously conducted by each agent hence partly relies on such a fundamental approach (via its cointegrated approximation of the fundamental value $\mathcal{T}^{j}(t)$), and partly on a chartist approach (via its forecasting of historical market data). In fact, based on the market state, each agent learns by reinforcement how much more fundamentalist or chartist it should be in its asset price estimation (see below). The entire simulation procedure follows four major steps, as shown below. 

i- \textit{Agents initialisation}: We consider a trading year on the London Stock Exchange as $T_y=281$ time steps, and similarly set a trading month and week as $T_m=21$ and $T_w=5$, respectively. At time $t=0$, all agents have reinforcement learning policies set with equal probabilities associated with each state-action pair. In other words, the agents do not yet know how to forecast nor trade: they need to engage in exploratory action selections. At the beginning of the simulation, they are given a learning phase of $1000$ time steps to learn this, after which their portfolio net asset value is set back to its initial value. Let $\mathcal{U} ()$ and $\mathcal{U} \{ \}$ denote the continuous and discrete uniform distributions, respectively. Then, at time $t=0$, all agents are initialised with the following parameters:
\begin{itemize}
\item A reinforcement learning rate $\alpha \sim \mathcal{U} (0.05, 0.20)$: This is proper to both reinforcement learning algorithms (see below). 
\item An investment horizon $\tau^{i} \sim \mathcal{U} \{T_w, 6T_m \}$: This is the number of time steps after which the agent liquidates its position.
\item A memory interval $h^{i} \sim \mathcal{U} \{T_w, T-\tau^{i}-2T_w \}$: This is the past lag interval used by each agent for its learning process.
\item A reflexivity parameter $\rho^{i} \sim \mathcal{U} (0, 100\%)$: This gauges how fundamentalist or chartist the agent is, via an average of its price forecast weighted by $P^{j}(t)$ and $B^{i,j}(t)$.
\item A drawdown limit $l^{i} \sim \mathcal{U} (50 \%, 60\%)$: This is the threshold of the year-to-date, peak-to-bottom loss in net asset value, above which the agent is listed as bankrupt, and unable to interact with the market anymore.
\item A trading window $w^{i} \sim \mathcal{U} \{T_w, \tau^{i} \}$: This assesses the optimal trading time for sending an order.
\item A transaction gesture $g^{i} \sim \mathcal{U} (0.2, 0.8)$: This scales with the bid-ask spread (see below) to set how far above or below the value of its own stock pricing the agent is willing to deal the transaction.
\end{itemize}

\textit{Order book}: The double-auction order book~\cite{Mota2016} receives the trading orders of the agents at each time step $t$ of the simulation in a random manner, so that no agent has any priority over another. The order book then sorts the bids from largest to lowest, and the offers from lowest to largest, for each stock $j$. After matching these bid and ask orders with one another, the order book then clears them when the ask prices are below the bid prices, starting from the top of the order book. Such transactions are cleared at mid-price between the two. If the quantity of stocks that one of the agent wanted to trade is not entirely transacted, this brings in another transaction with the next agent in queue. The market price $P^{j}(t+1)$ at the next time step of the simulation is defined as the price of the last transaction that was cleared by the order book at time $t$. Then the transaction volume $V^{j}(t+1)$ is taken as the total number of stocks that were transacted at time $t$, and the bid-ask spread $S^{j}(t+1)$ as the absolute difference between the average of all bids and asks involved in transactions cleared at time $t$. Based on current industrial broker fees for the London Stock Exchange~\cite{BrokerFees}, we model each agent's transaction as subject to broker fees of $0.1 \%$. Also, based on an approximation of the one-year gilt or UK bond yield average between January $2008$ and January $2018$~\cite{Gilt}, we model the risk-free assets of each agent as subject to an annual risk-free rate of $1.0 \%$. Finally, based on the current dividend impacts of FTSE 250 stocks~\cite{DividendYield}, we set the annual stock dividend yield of each agent to $2.0 \%$. In this Paper, we set $J=1$ and thus study agents trading only one stock, over the course of $S$ simulation runs. 

\begin{figure}[!htbp]
\begin{centering}
\includegraphics[scale=0.26]{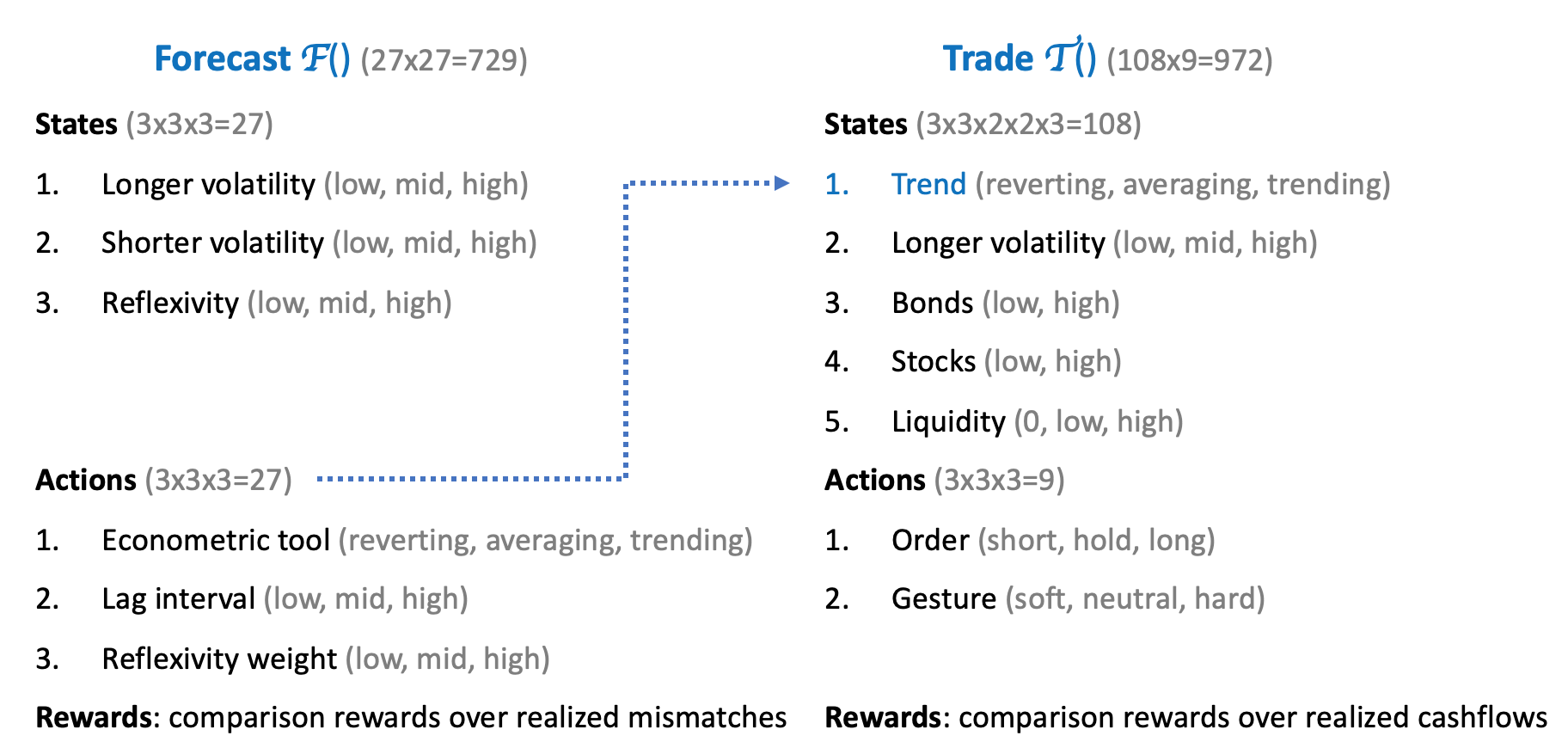}
\caption{\label{A1b} Procedure of the two reinforcement learning algorithms $\mathcal{F}()$ and $\mathcal{T}()$ autonomously performed by the agents, the latter taking as input the forecast of the former. Algorithm $\mathcal{F}()$ relies on a combination of $27$ states and $27$ actions, whereas $\mathcal{T}()$ on $108$ states and $9$ actions.}
\end{centering}
\end{figure}

\textit{Agents reinforcement learning}: As shown on Fig. \ref{A1b}, the agents rely on two distinct reinforcement learning algorithms before sending any transaction order to the order book~\cite{SuttonBarto,Wiering2012,Csaba2010,Lussange2019,Lussange2019b}. A first one $\mathcal{F}^{i}$, is aimed at providing accurate market price forecasting at the agent's proprietary investment horizon $\tau^{i}$, taking into consideration the market prices $P^{j}(t)$ together with the agent's own cointegrated approximation of the stock fundamental values $\mathcal{T}^{j}(t)$. This first algorithm thus forecasts a future price estimation that in turn serves as input to a second reinforcement algorithm $\mathcal{T}^{i}$, whose task is to decide which trading strategy should be employed at this time $t$, and hence which transaction order should be sent to the order book. In these two reinforcement algorithms, a direct policy search algorithm is conducted by each agent $i$, for each stock $j$, and at each time step $t$. According to the general reinforcement procedure, the states $s$ of the environment are monitored by each agent, which then selects a given action $a$ according to its current policy $\pi(s,a)=\mathbb{P}(s,a)$, eventually obtaining a resulting reward $r$ for this action, hence allowing the agent to update its policy accordingly for a more efficient action selection at next time steps, in an iterative way. 
\begin{itemize}
\item \textit{Forecasting algorithm} $\mathcal{F}^{i}$: The states of the environment are given by all the possible combinations of the combination of the longer-term volatility of the stock prices $s_0^{\mathcal{F}}=\{low, mid, high\}$, their shorter-term volatility $s_1^{\mathcal{F}}=\{low, mid, high\}$, and the gap between the agent's own present fundamental valuation and the market price $s_2^{\mathcal{F}}=\{low, mid, high\}$. With these states of the environment, the agent also has the following possible action selection: choosing a simple forecasting econometric tool based on mean-reverting, averaging, or trend-following market prices $a_0^{\mathcal{F}}=\{revert, mean, trend\}$, choosing the size of the historical lag interval for this forecast $a_1^{\mathcal{F}}=\{low, mid, high\}$, and choosing the weight of its own fundamental stock pricing in an overall future price estimation, that is both fundamentalist and chartist $a_2^{\mathcal{F}}=\{low, mid, high\}$. In this environment states, if these actions are taken at time $t$, the ensuing rewards are then retrospectively derived $\tau^{i}$ time steps later, as $r^{\mathcal{F}}=\{-4, -2, -1, 1, 2, 4 \}$, according to percentiles in the distribution of the agent's mismatches between past forecasts at time $t$ and their actual price realization at time $t+\tau^{i}$. Along with these rewards, the policy of $\mathcal{F}^{i}$ is then updated so as to increase the probabilities associated with the state-action pairs yielding lesser prediction errors. With $3 \times 3 \times 3 = 27$ states and $3 \times 3 \times 3 = 27$ actions, the first reinforcement learning algorithm $\mathcal{F}^{i}$ thus has $27 \times 27 = 729$ state-action pairs for its exploration and exploitation.

\item \textit{Trading algorithm} $\mathcal{T}^{i}$: The states of the environment are derived from a combination of the forecasting output of the former algorithm $s_0^{\mathcal{T}}=\{revert, mean, trend\}$, the longer-time price volatility $s_1^{\mathcal{T}}=\{low, mid, high\}$, the level of the agent's risk-free assets since the beginning of the simulation $s_2^{\mathcal{T}}=\{low, high\}$, the level of the agent's stock holdings since the beginning of the simulation $s_3^{\mathcal{T}}=\{low, high\}$, and the traded volumes of stock $j$ at former time step $s_4^{\mathcal{T}}=\{zero, low, high\}$. With these states of the environment, the agent also has the following possible action selection: sending a transaction order to the order book as holding, buying, or selling a position in a given amount proportional to its risk-free assets or stock holdings $a_0^{\mathcal{T}}=\{short, hold, long\}$, and at what price wrt. the current bid-ask spread $a_1^{\mathcal{T}}=\{soft, neutral, hard\}$. In these environment states, if these actions are taken at time $t$, the ensuing rewards are then retrospectively derived $\tau^{i}$ time steps later, as $r^{\mathcal{T}}=\{-4, -2, -1, 1, 2, 4 \}$, according to percentiles in the distribution of the agent's past cashflow differences. Along with these rewards, the policy of $\mathcal{T}^{i}$ is then likewise updated so as to increase the probabilities of action-pairs yielding lesser prediction error. With $3 \times 3 \times 2 \times 2 \times 3 = 108$ states and $3 \times 3 = 9$ actions, the first reinforcement learning algorithm $\mathcal{F}^{i}$ thus has $108 \times 9 = 972$ state-action pairs for exploration and exploitation.
\end{itemize}

\section{Impact of tick size}
\label{SectionIII}

\begin{figure}[!htbp]
\begin{centering}
\includegraphics[scale=0.53]{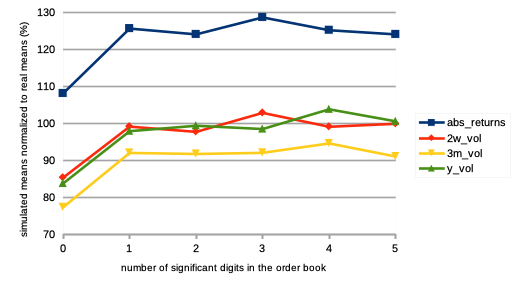}
\caption{\label{O1} Means of absolute logarithmic returns (blue) and volatilities (defined as standard deviations of price normalised to price itself $\sigma/P(t)$) computed over lags of two weeks (red), three months (yellow), and one year (green) intervals, for simulations with an order book accounting for transaction prices with $0, 1, 2, 3, 4$ or $5$ significant digits only. These values are displayed as percentages of those coming from real data. The simulations are generated with parameters $I=500$, $T=2875$, $S=20$.}
\end{centering}
\end{figure}

\begin{figure}[!htbp]
\begin{centering}
\includegraphics[scale=0.53]{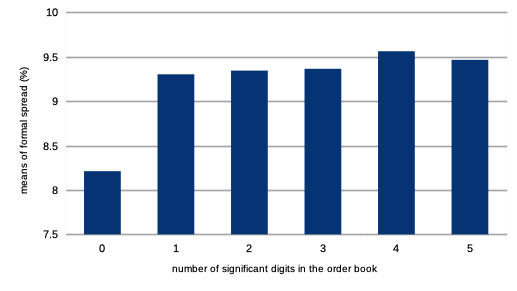}
\caption{\label{O2a} Means of bid-ask spreads $S^{j}(t)$ (as a percentage of the price, and defined as the absolute average of all bid and ask prices of the orders leading to transactions), for simulations with an order book accounting for transaction prices with $0, 1, 2, 3, 4$ or $5$ significant digits only. The simulations are generated with parameters $I=500$, $T=2875$, $S=20$.}
\end{centering}
\end{figure}

\begin{figure}[!htbp]
\begin{centering}
\includegraphics[scale=0.53]{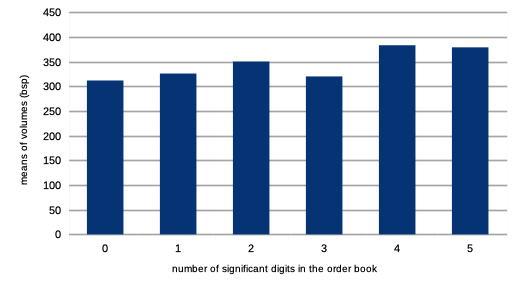}
\caption{\label{O3} Means of trading volumes $V^{j}(t)$ (as basis points of the number of stock outstanding), for simulations with an order book accounting for transaction prices with $0, 1, 2, 3, 4$ or $5$ significant digits only. The simulations are generated with parameters $I=500$, $T=2875$, $S=20$.}
\end{centering}
\end{figure}

\begin{figure}[!htbp]
\begin{centering}
\includegraphics[scale=0.53]{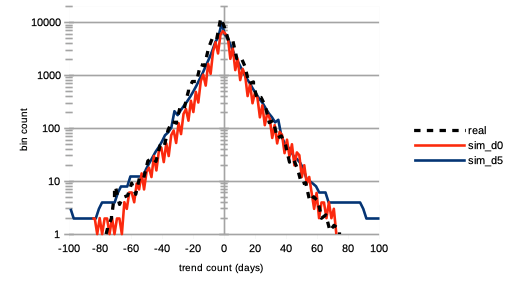}
\caption{\label{O2b} Distribution of the number of consecutive days of rising prices (positive values) and dropping prices (negative values) for both real data (dashed black curve) and simulated data with a tick size of $0$ (red continuous curve) and $5$ (blue continuous curve) significant digits. The simulations are generated with parameters $I=500$, $T=2875$, $S=20$.}
\end{centering}
\end{figure}

\begin{figure}[!htbp]
\begin{centering}
\includegraphics[scale=0.53]{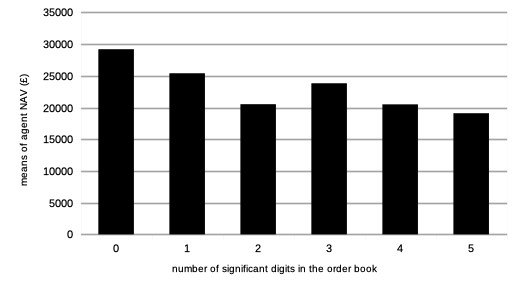}
\caption{\label{O4} Means of the net asset value of all agents at $t=T$, for simulations with an order book accounting for transaction prices with $0, 1, 2, 3, 4$ or $5$ significant digits only. The simulations are generated with parameters $I=500$, $T=2875$, $S=20$.}
\end{centering}
\end{figure}

Such a groundwork and MAS stock market model can be used to quantitatively study the market impact of an ever-lower order book tick size, which is the minimum regulatory interval between prices in an order book. With regards to this, one can especially mention the live experiment called the \textit{Tick Size Pilot Program}~\cite{Hu2018,FINRA2018} conducted from $2016$ to $2018$ by the U. S. Securities and Exchange Commission (SEC). This experiment examined the effects on average stock market quality of a smaller order book tick size, albeit only for the stocks of smaller capitalisation companies. Over all, this Program found a deterioration in market quality, with ensuing larger spreads and price volatilities. Similar recent studies have been conducted in Europe~\cite{AMF2018}, derived from the impact of the \textit{Markets in Financial Instruments Directive} (MiFID 2), with similar conclusions. In our MAS approach, we model the order book tick size by the number of possible significant digits of the transaction prices, starting with prices set as integers (or zero significant digits) in order to identify general trends as we vary tick sizes.  
\begin{itemize}
\item[--] We see on Fig. \ref{O1} an increase and quick convergence in price absolute returns and volatilities at several time intervals, for smaller tick sizes (or equivalently, larger significant digits). Notice this quantities are normalised, as divided by real data metrics and expressed in percent. 
\item[--] Likewise, we similarly see on Fig. \ref{O2a} an increase and convergence in the average formal market bid-ask spreads. Under our model assumptions, this would tend to confirm the results of the other aforementioned studies~\cite{Hu2018,AMF2018}. Notice this is counter-intuitive, as one could have expected larger tick sizes to constrain the agent bid and ask prices and thus widen the spread. 
\item[--] As seen on Fig. \ref{O3}, we find a mild increase in trading volumes, in line with the results of previous studies such as~\cite{AMF2018,Chou2006}, the latter finding that ETFs experienced an increase in trading volume following decimalisation. 
\item[--] If we study the market regimes by recording the number of consecutive days of daily increasing prices (positive values) and decreasing prices (negative values), as shown on the distribution of Fig. \ref{O2b}, we find that smaller tick sizes greatly amplify both bullish and bearish market regimes, as one can see from the tails of the distributions. This is congruent with the increasing price volatilities and bid-ask spreads mentioned above. We posit this to be a strong factor of market instability. Also, we find no correlation between the average number of market crashes and varying tick sizes: over all significant digits, we find an average of $1.86 \pm 0.61$ crash per simulation run.
\item[--] At the agents level, we see on Fig. \ref{O4} that decreasing the tick size tends to lower the general wealth of the agents on the long-term. This is seen through a steady decrease in average net asset value of all agents at the end of each simulation. The market quality wrt. overall agent economic growth and survival rates is thus not gaining from smaller tick sizes. 

\end{itemize}

Thus, we find overall a greater market instability due to smaller tick sizes allowed in the order book, in line with the SEC's pilot program mentioned above~\cite{Hu2018,FINRA2018}. From the perspective of agent economic growth and survival rates, we find such lower tick sizes to also diminish overall market quality.

\clearpage

\section{Impact of metaorders}
\label{SectionIV}

Following a rich literature on the topic of metaorders~\cite{Bucci2019,Said2019}, we want to study the impact of very large orders on the market microstructure. In order to do this, we randomly select an agent once every trading year of the simulations according to a uniform distribution, and endow it with an arbitrary large amount of stock holdings (in case of a short order sent by the agent to the order book) or risk-free assets (in case of a buy order). As soon as this agent has sent its very large order to the order book, its portfolio's net asset value is reset to its former value. Not unlike~\cite{Bucci2019}, we then record for a given stock $j$ the following statistics: 
\begin{eqnarray}
\rho(t)=\frac{Q(t)}{Q_{tot}}
\label{eq1}
\end{eqnarray}
\begin{eqnarray}
\mathcal I^P(t) = \frac{\sigma^P_{[t \rightarrow t+\tau]} - \sigma^P_{[t - \tau \rightarrow t]}}{\sigma^P_{[t - \tau \rightarrow t}]}
\label{eq2}
\end{eqnarray}

\noindent Here, $\rho(t)$ is the ratio of traded stocks at time $t$ expressed in percent, $Q(t)$ is the number of stocks traded at time $t$, and $Q_{tot}$ the total shares outstanding. $I^P(t)$ is the impact on price volatilities at time $t$. We denote $\sigma^P_{[t_0 \rightarrow t_1]}$ as the standard deviation of prices $P$ on the interval $t \in [t_0, t_1]$. The quantity $\tau$ is a given time interval. 

\vspace{1mm}

\begin{figure}[!htbp]
\begin{centering}
\includegraphics[scale=0.53]{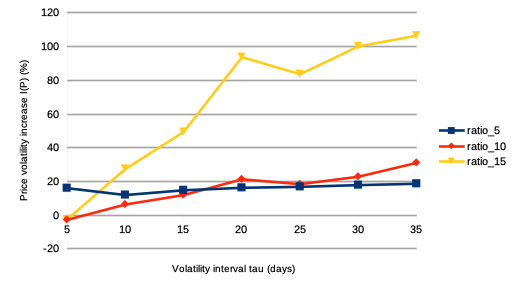}
\caption{\label{M2} Average impacts on price volatility $I^P(t)$ of orders of ratio $\rho(t) \in [0, 5\%]$ (blue curve), $\rho(t) \in [5, 10\%]$ (red curve), and $\rho(t) \in [10, 15\%]$ (yellow curve), as a function of such volatility interval $\tau$. The simulations are generated with parameters $I=500$, $T=2875$, $S=150$.}
\end{centering}
\end{figure}

On Fig. \ref{M2}, we then show the average impacts on price volatility $I^P(t)$, given by equ. \ref{eq2}, for different sizes of orders: those with ratios $\rho(t) \in [0, 5\%]$ (blue curves), $\rho(t) \in [5, 10\%]$ (red curves), and $\rho(t) \in [10, 15\%]$ (yellow curves). Both results are displayed as functions of various volatility intervals $\tau$. 

We find that price volatilities display a sharp increase, especially for larger intervals $\tau$ and orders of larger amplitudes $\rho$. We posit this increase with larger intervals $\tau$ to be caused by the greater amount of total assets and cashflow being present among agents after the very large order has been cleared by the order book. In other words, a continuous market activity proceeds from this very large order, thereby increasing the impact $I^P$ over longer time intervals. This is why we see steeper slopes for the curves corresponding to larger orders. 

Another thing one can notice from Fig. \ref{M2}, is that irrespectively of $\tau$, the impact $I^P$ corresponding to very large orders seems to scale non-linearly with $\rho(t)$. In other words, the values of $I^P$ seem to increase exponentially with $\rho(t)$, and very large orders, above a certain threshold of cashflow, may cause irreversible changes to a market.  

Finally, as one could have expected, all these curves display a positive increase in price volatility $I^P(t)$. As a conclusion, we see that increasing the order size a strong impact on the volatilities of prices, and that extra-large orders can make volatility explode and hence deteriorate overall market quality.

\clearpage

\section{Impact of trading frequency}
\label{SectionV}

\begin{figure}[!htbp]
\begin{centering}
\includegraphics[scale=0.53]{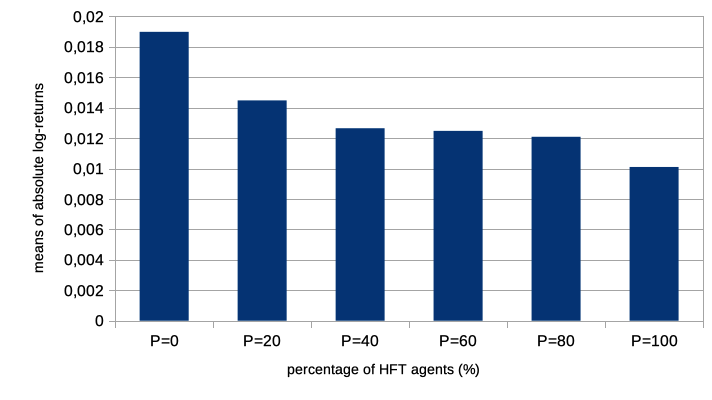}
\caption{\label{B1} Means of all absolute logarithmic price returns for simulations with a percentage $p$ of agents corresponding to $p=0\%, 20\%, 40\%, 60\%, 80\%, 100\%$ of the total agent population, that are initialized at time $t=0$ with an investment horizon $\tau^{i} \sim \mathcal{U} [T_w, 2T_w($, while the remaining $100-p \%$ agents' investment horizon $\tau^{i}$ are drawn from $\mathcal{U} \{T_w, 6T_m \}$. The simulations are generated with parameters $I=500$, $T=2875$ (corresponding to about $11$ years), and $S=20$.}
\end{centering}
\end{figure}

\begin{figure}[!htbp]
\begin{centering}
\includegraphics[scale=0.53]{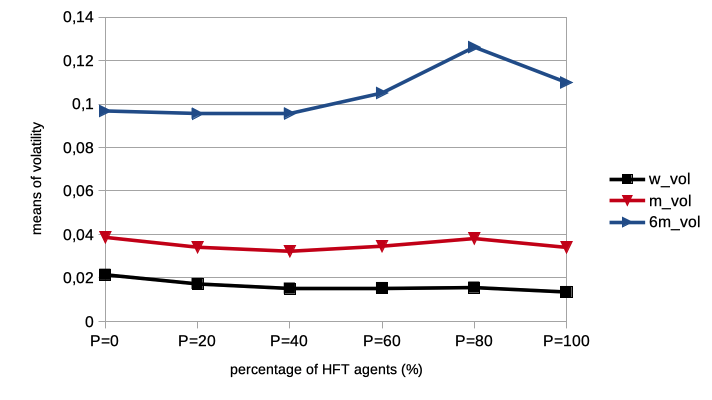}
\caption{\label{B2} Means of all volatilities (defined as standard deviations of price normalised to price itself $\sigma/P(t)$) computed over lags of one weeks (black), one month (red), and six months (blue) intervals, for simulations with a percentage $p$ of agents corresponding to $p=0\%, 20\%, 40\%, 60\%, 80\%, 100\%$ of the total agent population, that are initialized at time $t=0$ with an investment horizon $\tau^{i} \sim \mathcal{U} [T_w, 2T_w($, while the remaining $100-p \%$ agents' investment horizon $\tau^{i}$ are drawn from $\mathcal{U} \{T_w, 6T_m \}$. The simulations are generated with parameters $I=500$, $T=2875$ (corresponding to about $11$ years), and $S=20$.}
\end{centering}
\end{figure}

\begin{figure}[!htbp]
\begin{centering}
\includegraphics[scale=0.53]{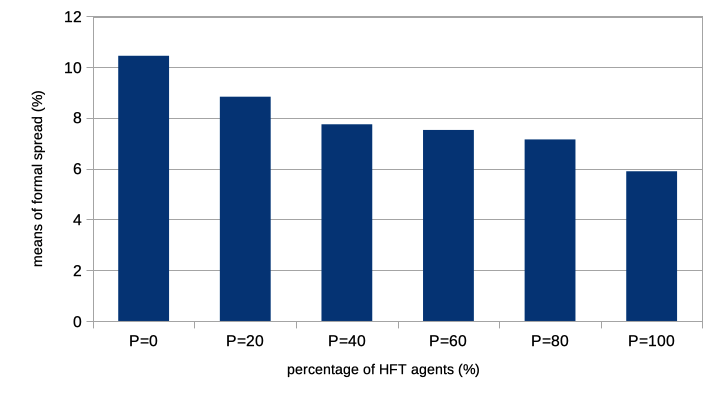}
\caption{\label{B4} Means of all bid-ask spread in percent of price, for a percentage $p$ of agents corresponding to $p=0\%, 20\%, 40\%, 60\%, 80\%, 100\%$ of the total agent population, that are initialized at time $t=0$ with an investment horizon $\tau^{i} \sim \mathcal{U} [T_w, 2T_w($, while the remaining $100-p \%$ agents' investment horizon $\tau^{i}$ are drawn from $\mathcal{U} \{T_w, 6T_m \}$. The simulations are generated with parameters $I=500$, $T=2875$ (corresponding to about $11$ years), and $S=20$.}
\end{centering}
\end{figure}

\begin{figure}[!htbp]
\begin{centering}
\includegraphics[scale=0.53]{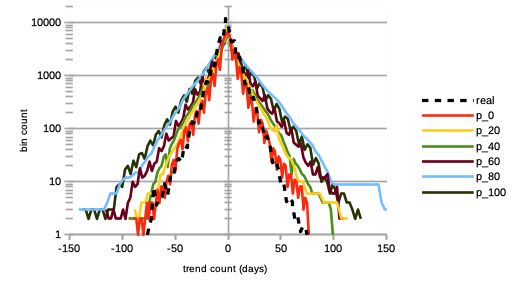}
\caption{\label{B5} Distribution of the number of consecutive days of rising prices (positive values) and dropping prices (negative values). This is for both real (dashed black curve) and simulated (continuous curves) data, the latter being for a percentage $p$ of agents corresponding to $p=0\%$ (red), $p=20\%$ (yellow), $p=40\%$ (green), $p=60\%$ (brown), $p=80\%$ (light blue), and $p=100\%$ (dark green) of the total agent population, that are initialized at time $t=0$ with an investment horizon $\tau^{i} \sim \mathcal{U} [T_w, 2T_w($, while the remaining $100-p \%$ agents' investment horizon $\tau^{i}$ are drawn from $\mathcal{U} \{T_w, 6T_m \}$. The simulations are generated with parameters $I=500$, $T=2875$ (corresponding to about $11$ years), and $S=20$.}\end{centering}
\end{figure}

We now want to study the market impact of increasingly larger percentages of agents set with a higher frequency of trading. The interest of such a study can be linked with the role played by high frequency trading in modern stock exchanges. However, we recall that our model is calibrated to daily stock market data, and that the interest of such a study is simply to gauge the effect of larger populations of agents trading more frequently and at shorter time-scales within the scope of our model assumptions. As increasingly more agents trade at ever higher frequencies, one could ask if this is beneficiary or not to market quality~\cite{Angel2014,Breckenfelder2019}. The rush for higher frequency trading can be linked to the importance of reducing portfolio drawdowns and smoothening equity curves in the asset management industry. We here model high frequency trading by agents being initialised with a much lower investment horizon $\tau^{i}$, now drawn from a discrete uniform distribution $\mathcal{U} [T_w, 2T_w($. Compared to the other agents, these high frequency agents thus have a trading horizon reduced by one order of magnitude, on average. For simulations with increasing percentages of such agents, we observe the following:
\begin{itemize}
\item[--] We see on Fig. \ref{B1} a strong decrease in absolute logarithmic price returns. This can be explained from the order book, at each time step $t$ being filled with a larger quantity of transaction orders, and hence invariably, a greater probability for spread-centered orders to be sent, matched, and cleared by the order book. 
\item[--] We see on Fig. \ref{B2} a steady decrease in short-term price volatility, together with an increase in long-term volatility. This again could be explained by the greater quantity of trading orders being cleared by the order book at each time step, smoothening the stock price curve at shorter scales. At larger scales, we find increasing volatility (here at the semester-interval), in line with longer market regimes (see below). 
\item[--] We see on Fig. \ref{B3} a very strong increase in trading volumes, as expected, since agents with a shorter investment horizon will tend to trade more frequently. 
\item[--] We see on Fig. \ref{B4} a steady decrease in market bid-ask spread, which can be explained by the larger amounts of trading orders being sent to the order book. 
\item[--] We see on Fig. \ref{B5} a greater propensity for longer bull and bear market regimes. Furthermore, we find that the rates of agent bankruptcies remain stable, regardless of these varying percentages, with an average of $2.86 \pm 0.97$ crashes (defined as a $20\%$ decrease in stock daily price) for each of the $S=20$ simulation runs (and each run lasting $T=2875$ time steps). 
\end{itemize}

Under our model assumptions, increasing numbers of such higher frequency trading agents are thus posited to be beneficial to stock market stability, by lowering general volatility and increasing trading volumes, and thus tackling the issue of market illiquidity propitious to crashes. Although the market impact of high frequency trading has been an active field of research in quantitative finance~\cite{Breckenfelder2019}, there is, to our knowledge, no other study that probes the effects of larger \textit{populations} of high frequency traders in a given market. One should consider these results together with the means that have been proposed for high-frequency trading regulation~\cite{Currie2017}. Finally, one should consider and study from an economic perspective why larger proportions of higher frequency agents would bring larger amplitudes of both bearish and bullish market regimes.

\begin{figure}[!htbp]
\begin{centering}
\includegraphics[scale=0.53]{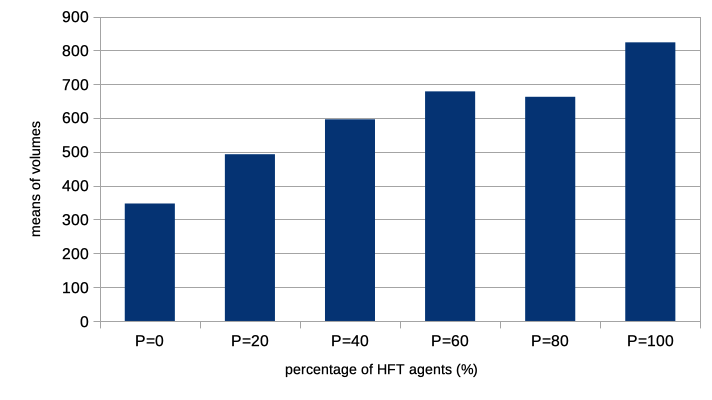}
\caption{\label{B3} Means of all trading volumes, for a percentage $p$ of agents corresponding to $p=0\%, 20\%, 40\%, 60\%, 80\%, 100\%$ of the total agent population, that are initialized at time $t=0$ with an investment horizon $\tau^{i} \sim \mathcal{U} [T_w, 2T_w($, while the remaining $100-p \%$ agents' investment horizon $\tau^{i}$ are drawn from $\mathcal{U} \{T_w, 6T_m \}$. The simulations are generated with parameters $I=500$, $T=2875$ (corresponding to about $11$ years), and $S=20$.}
\end{centering}
\end{figure}

\clearpage 

\section{Conclusion}
\label{SectionVI}
Based on a previous groundwork~\cite{Lussange2019,Lussange2019b} detailing an ABM stock market simulator where the agents autonomously learn to forecast and trade by reinforcement learning, we presented here some of its predictions concerning certain specific regulatory issues. We first studied the theoretical impact of lowered tick sizes in stock market order books, and found under our model's assumptions a greater market instability due to larger tick sizes allowed in the order book, in line with the SEC's pilot program mentioned above~\cite{Hu2018,FINRA2018}. From the perspective of agent economic growth and survival rates, we find such larger tick sizes to also diminish overall market quality. We then studied the impact of larger metaorders on the market, and found an increase in price volatility after a very large order has been cleared in the order book, scaling with the time interval of this volatility. Also we found this increase to be non-linear with the order size: it seems very large orders above a certain threshold can cause lasting and irreversible damage to the market volatility. This should be further studied, in a coming work. Finally, we studied the impact of larger populations of agents in a given market engaging in higher frequency trading, by modelling such agents with much shorter investment horizons. We found that larger populations of such agents are beneficial to stock market stability, lowering general volatility and increasing trading volumes, thereby tackling the issue of market illiquidity, that is propitious to crashes. A next improvement and field of study would be to calibrate our model to intraday data, and consider a time step to be equivalent, say, to a minute. This would allow us to check on market microstructure effects, and see how these would relate to the former regulatory trends.

\clearpage

\bibliographystyle{spmpsci}   
\bibliography{template}   

\begin{thebibliography}{10}
\providecommand{\url}[1]{{#1}}
\providecommand{\urlprefix}{URL }
\expandafter\ifx\csname urlstyle\endcsname\relax
  \providecommand{\doi}[1]{DOI~\discretionary{}{}{}#1}\else
  \providecommand{\doi}{DOI~\discretionary{}{}{}\begingroup
  \urlstyle{rm}\Url}\fi

\bibitem{DividendYield}
Current dividend impacts of {FTSE}-250 stocks.
\newblock \urlprefix\url{https://www.dividenddata.co.uk}.
\newblock Accessed: 2020-05-19

\bibitem{BrokerFees}
{IG} fees of {C}ontracts {F}or {D}ifference.
\newblock \urlprefix\url{https://www.ig.com}.
\newblock Accessed: 2020-05-19

\bibitem{Gilt}
{UK} one-year gilt reference prices.
\newblock \urlprefix\url{https://www.dmo.gov.uk}.
\newblock Accessed: 2020-05-19

\bibitem{AMF2018}
MIFID II: Impact of the New Tick Size Regime.
\newblock Autorite des Marches Financiers (2018)

\bibitem{Aloud2014}
Aloud, M.: Agent-based simulation in finance: design and choices.
\newblock Proceedings in Finance and Risk Perspectives ‘14  (2014)

\bibitem{Angel2014}
Angel, J.: When finance meets physics: The impact of the speed of light on
  financial markets and their regulation.
\newblock Wiley Online Library (2014)

\bibitem{FINRA2018}
Authority, F.I.R.: Assessment of the Plan to Implement a Tick Size Pilot
  Program.
\newblock FINRA (2018)

\bibitem{Benzaquen2018}
Benzaquen, M., Bouchaud, J.P.: A fractional reaction–diffusion description of
  supply and demand.
\newblock The European Physical Journal B \textbf{91(23)} (2018)

\bibitem{Bera2015}
Bera, A.K., Ivliev, S., Lillo, F.: Financial Econometrics and Empirical Market
  Microstructure.
\newblock Springer (2015)

\bibitem{Biondo2019}
Biondo, A.E.: Order book modeling and financial stability.
\newblock Journal of Economic Interaction and Coordination \textbf{14(3)}
  (2019)

\bibitem{Boero2015}
Boero, R., Morini, M., Sonnessa, M., Terna, P.: Agent-based models of the
  economy, from theories to applications.
\newblock Palgrave Macmillan (2015)

\bibitem{Bouchaud2018}
Bouchaud, J.P.: Handbook of Computational Economics \textbf{4} (2018)

\bibitem{Bouchaud2019}
Bouchaud, J.P.: Econophysics: Still fringe after 30 years?
\newblock arXiv:1901.03691  (2019)

\bibitem{Breckenfelder2019}
Breckenfelder, J.: Competition among high-frequency traders, and market
  quality.
\newblock European Central Bank Working Paper Series nr. 2290 (2019)

\bibitem{Bucci2019}
Bucci, F., Benzaquen, M., Lillo, F., Bouchaud, J.P.: Slow decay of impact in
  equity markets: insights from the ancerno database.
\newblock arXiv:1901.05332  (2019)

\bibitem{Charpentier2020}
Charpentier, A., Elie, R., Remlinger, C.: Reinforcement learning in economics
  and finance.
\newblock arXiv:2003.10014  (2020)

\bibitem{Chiarella2007}
Chiarella, C., Iori, G., Perell, J.: The impact of heterogeneous trading rules
  on the limit order book and order flows.
\newblock arXiv:0711.3581  (2007)

\bibitem{Chou2006}
Chou, R.K., Chung, H.: Decimalization, trading costs, and information
  transmission between etfs and index futures.
\newblock Journal of Futures Markets \textbf{26(2)}, 131--151 (2006)

\bibitem{Cont2005}
Cont, R.: Chapter 7 - Agent-Based Models for Market Impact and Volatility.
\newblock A Kirman and G Teyssiere: Long memory in economics, Springer (2005)

\bibitem{Cristelli2014}
Cristelli, M.: Complexity in Financial Markets.
\newblock Springer (2014)

\bibitem{Currie2017}
Currie, W.L., Seddon, J.J.M.: The regulatory, technology and market ?dark arts
  trilogy? of high frequency trading: a research agenda.
\newblock Journal of Information Technology \textbf{32} (2017)

\bibitem{Dayan2008}
Dayan, P., Daw, N.D.: Decision theory, reinforcement learning, and the brain.
\newblock Cognitive, Affective, and Behavioral Neuroscience \textbf{8(4)},
  429--453 (2008)

\bibitem{Deng2017}
Deng, Y., Bao, F., Kong, Y., Ren, Z., Dai, Q.: Deep direct reinforcement
  learning for financial signal representation and trading.
\newblock IEEE Trans. on Neural Networks and Learning Systems \textbf{28(3)}
  (2017)

\bibitem{Eickhoff2018}
Eickhoff, S.B., Yeo, B.T.T., Genon, S.: Imaging-based parcellations of the
  human brain.
\newblock Nature Reviews Neuroscience \textbf{19}, 672--686 (2018)

\bibitem{ErevRoth2014}
Erev, I., E.Roth, A.: Maximization, learning and economic behaviour.
\newblock PNAS \textbf{111}, 10818--10825 (2014)

\bibitem{Fama1970}
Fama, E.: Efficient capital markets: A review of theory and empirical work.
\newblock Journal of Finance \textbf{25}, 383--417 (1970)

\bibitem{Franke2011}
Franke, R., Westerhoff, F.: Structural stochastic volatility in asset pricing
  dynamics: Estimation and model contest.
\newblock BERG Working Paper Series on Government and Growth \textbf{78} (2011)

\bibitem{Frydman2016}
Frydman, C., Camerer, C.F.: The psychology and neuroscience of financial
  decision making.
\newblock Trends in Cognitive Sciences \textbf{20}, 661--675 (2016)

\bibitem{Furtado2016}
Furtado, B.A., Eberhardt, I.D.R.: A simple agent-based spatial model of the
  economy: Tools for policy.
\newblock Journal of Artificial Societies and Social Simulation \textbf{19(4)}
  (2016)

\bibitem{Ganesh2019}
Ganesh, S., Vadori, N., Xu, M., Zheng, H., Reddy, P., Veloso, M.: Reinforcement
  learning for market making in a multi-agent dealer market.
\newblock arXiv:1911.05892  (2019)

\bibitem{Gualdi2015}
Gualdi, S., Tarzia, M., Zamponi, F., Bouchaud, J.P.: Tipping points in
  macroeconomic agent-based models.
\newblock Journal of Economic Dynamics and Control \textbf{50}, 29--61 (2015)

\bibitem{Hu2018}
Hu, E., Hughes, P., Ritter, J., Vegella, P., Zhang, H.: U. S. Securities and
  Exchange Commission white papers  (2018)

\bibitem{Hu2019}
Hu, Y.J., Lin, S.J.: Deep reinforcement learning for optimizing portfolio
  management.
\newblock 2019 Amity International Conference on Artificial Intelligence
  (2019)

\bibitem{Huang2015}
Huang, W., Lehalle, C.A., Rosenbaum, M.: Simulating and analyzing order book
  data: the queue-reactive model.
\newblock Journal of the American Statistical Association \textbf{110}, 509
  (2015)

\bibitem{Lefebvre2017}
Lefebvre, G., Lebreton, M., Meyniel, F., Bourgeois-Gironde, S., Palminteri, S.:
  Behavioural and neural characterization of optimistic reinforcement learning.
\newblock Nature Human Behaviour \textbf{1(4)} (2017)

\bibitem{Lipski2013}
Lipski, J., Kutner, R.: Agent-based stock market model with endogenous
  agents’ impact.
\newblock arXiv:1310.0762  (2013)

\bibitem{Lussange2018}
Lussange, J., Belianin, A., Gutkin, B., Bourgeois-Gironde, S.: Learning and
  cognition in financial markets: A paradigm shift for agent-based models.
\newblock Proceedings of SAI Intelligent Systems Conference pp. 241--255 (2020)

\bibitem{Lussange2019}
Lussange, J., Lazarevich, I., Bourgeois-Gironde, S., Palminteri, S., Gutkin,
  B.: Modelling stock markets by multi-agent reinforcement learning.
\newblock Computational Economics pp. 1--35 (2020)

\bibitem{Lussange2019b}
Lussange, J., Palminteri, S., Bourgeois-Gironde, S., Gutkin, B.: Stock price
  formation: useful insights from a multi-agent reinforcement learning model.
\newblock arXiv:1910.05137  (2020)

\bibitem{Murray1994}
Murray, M.P.: A drunk and her dog: An illustration of cointegration and error
  correction.
\newblock The American Statistician \textbf{48(1)}, 37--39 (1994)

\bibitem{Mota2016}
N, R.M., Larralde, H.: A detailed heterogeneous agent model for a single asset
  financial market with trading via an order book.
\newblock arXiv:1601.00229  (2016)

\bibitem{Neuneier1997}
Neuneier, R.: Enhancing q-learning for optimal asset allocation.
\newblock Proc. of the 10th International Conference on Neural Information
  Processing Systems  (1997)

\bibitem{Palminteri2015}
Palminteri, S., Khamassi, M., Joffily, M., Coricelli, G.: Contextual modulation
  of value signals in reward and punishment learning.
\newblock Nature communications pp. 1--14 (2015)

\bibitem{Said2019}
Said, E., Ayed, A.B.H., Husson, A., Abergel, F.: Market impact: A systematic
  study of limit orders.
\newblock arXiv:1802.08502  (2019)

\bibitem{Silver2018}
Silver, D., Hubert, T., Schrittwieser, J., Antonoglou, I., Lai, M., Guez, A.,
  Lanctot, M., Sifre, L., Kumaran, D., Graepel, T., Lillicrap, T., Simonyan,
  K., Hassabis, D.: A general reinforcement learning algorithm that masters
  chess, shogi and go through self-play.
\newblock Science \textbf{362}(6419), 1140--1144 (2018)

\bibitem{Sirignano2019}
Sirignano, J., Cont, R.: Universal features of price formation in financial
  markets: perspectives from deep learning.
\newblock Quantitative Finance \textbf{19(9)} (2019)

\bibitem{Sornette2014}
Sornette, D.: Physics and financial economics (1776-2014): Puzzles, ising and
  agent-based models.
\newblock Rep. Prog. Phys. \textbf{77} (2014)

\bibitem{Spooner2018}
Spooner, T., Fearnley, J., Savani, R., Koukorinis, A.: Market making via
  reinforcement learning.
\newblock Proceedings of the 17th AAMAS  (2018)

\bibitem{SuttonBarto}
Sutton, R., Barto, A.: Reinforcement Learning, second edition: An Introduction.
\newblock Bradford Books (2018)

\bibitem{Csaba2010}
Szepesvari, C.: Algorithms for Reinforcement Learning.
\newblock Morgan and Claypool Publishers (2010)

\bibitem{Wah2013}
Way, E., Wellman, M.P.: Latency arbitrage, market fragmentation, and
  efficiency: a two-market model.
\newblock Proceedings of the fourteenth ACM conference on Electronic commerce
  pp. 855--872 (2013)

\bibitem{Westerhoff2008}
Westerhoff, F.H.: The use of agent-based financial market models to test the
  effectiveness of regulatory policies.
\newblock Jahrbucher Fur Nationalokonomie Und Statistik \textbf{228(2)}, 195
  (2008)

\bibitem{Wiering2012}
Wiering, M., van Otterlo, M.: Reinforcement Learning: State-of-the-Art.
\newblock Springer, Berlin, Heidelberg (2012)

\end{thebibliography}

\clearpage

\section{Supplementary material}
\label{SectionVII}

\begin{figure}[!htbp]
\begin{centering}
\includegraphics[scale=0.53]{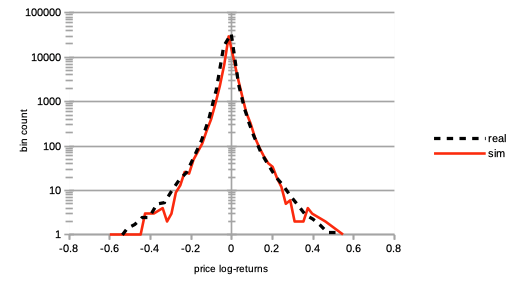}
\caption{\label{S1} Distribution of logarithmic returns of prices $\log [P(t)/P(t-1)]$ of real (dashed black curve) and simulated (continuous red curve) data. The simulations are generated with parameters $I=500$, $T=2875$, and $S=20$.}
\end{centering}
\end{figure}

\begin{figure}[!htbp]
\begin{centering}
\includegraphics[scale=0.53]{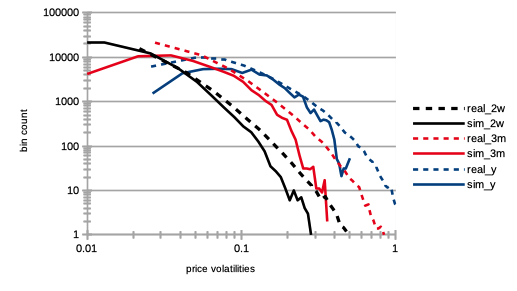}
\caption{\label{S2} Distribution of volatilities (defined as standard deviations of price normalised to price itself $\sigma/P(t)$) computed over lags of two weeks (black), three months (red), and one year (blue) intervals for both real (dashed curves) and simulated (continuous curves) data. The simulations are generated with parameters $I=500$, $T=2875$, and $S=20$.}
\end{centering}
\end{figure}

\begin{figure}[!htbp]
\begin{centering}
\includegraphics[scale=0.53]{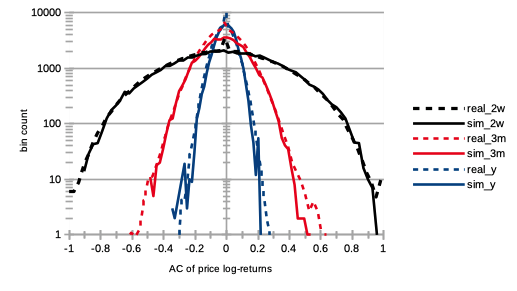}
\caption{\label{S3} Distribution of autocorrelations of the logarithmic returns of prices at each time step $t$ between intervals $[t-\Delta, t]$ and $[t-2\Delta, t-\Delta]$, over lags $\Delta$ of two weeks (black), three months (red), and one year (blue) intervals for both real (dashed curves) and simulated (continuous curves) data. The simulations are generated with parameters $I=500$, $T=2875$, and $S=20$.}
\end{centering}
\end{figure}

\begin{figure}[!htbp]
\begin{centering}
\includegraphics[scale=0.53]{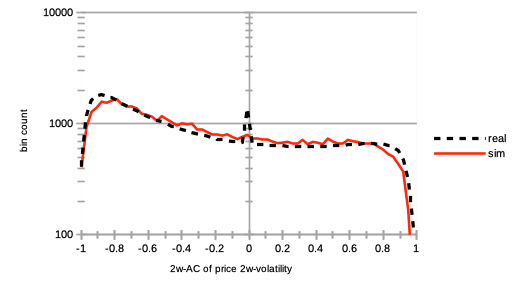}
\caption{\label{S4} Distribution of autocorrelations of two weeks-interval volatilities at each time step $t$ between intervals $[t-\Delta, t]$ and $[t-2\Delta, t-\Delta]$ for $\Delta=2T_{w}$, for both real (dashed black curve) and simulated (continuous red curve) data. The simulations are generated with parameters $I=500$, $T=2875$, and $S=20$.}
\end{centering}
\end{figure}

We recall here on Fig. \ref{S1}-\ref{S4} the main calibration results found in~\cite{Lussange2019}, pertaining to logarithmic price returns, price volatilities at different time scales, and auto-correlations of such logarithmic price returns and price volatilities. Via a careful parameter selection, we recall that such a model was calibrated to end-of-day prices and volumes of $642$ stocks traded on the London Stock Exchange, between the years $2007$ and $2018$.

\end{document}